\documentclass[11pt,a4paper]{article}
\usepackage{epsfig}
\newcommand{\I}{{\mathbf I}}
\newcommand{\R}{{\mathbf R}}
\renewcommand{\S}{{\mathbf S}}

\begin{document}

\begin{titlepage}

\hbox{}\vspace{1in}
\begin{center}

\textsc{ \Large
 An implementation of the Deutsch-Jozsa algorithm  on
a three-qubit NMR quantum computer }

\vspace{1in}

\textbf{Noah Linden}\footnote{\tt nl101@newton.cam.ac.uk}

Isaac Newton Institute for Mathematical
Sciences, 20 Clarkson Road, Cambridge, CB3 0EH, UK

and

Department of Applied Mathematics and Theoretical Physics, Silver Street,
Cambridge CB3 9EW, UK

\bigskip

\textbf{Herv\'e Barjat\footnote{\tt hb232@cus.cam.ac.uk} and
 Ray Freeman\footnote{{\tt rf110@cus.cam.ac.uk}; {\rm author to whom
correspondence should be addressed.}}}

Department of Chemistry, Lensfield Rd,
Cambridge CB2 1EW, UK

\vspace{1in}

\bigskip

%(PACS codes: ???)

\end{center}

\begin{center} \textsc{\large Abstract} \end{center} \medskip

A new approach to the implementation of a quantum computer by
high-resolution nuclear magnetic resonance (NMR) is described.  The key
feature is that two or more line-selective radio-frequency
pulses are applied simultaneously.   A three-qubit quantum computer has been
investigated using the 400 MHz NMR spectrum of the three coupled protons in
2,3-dibromopropanoic acid.  It has been employed to implement the
Deutsch-Jozsa algorithm for distinguishing between  constant  and
balanced functions.  The extension to
systems containing more coupled spins is straightforward and does not
require a more protracted experiment.

\end{titlepage}

{\openup 5pt

\section{Introduction}
 While there has long been theoretical interest in the notion of a quantum
computer, it was the series of recent results leading to the remarkable
algorithm  of Shor \cite{Shor94} for finding prime factors in polynomial time
which  led to the recent explosion of interest in the subject.  These
theoretical results have led many groups to try to realise a quantum computer
experimentally.  Nuclear magnetic resonance offers a  particularly attractive
implementation  of
quantum computers because nuclear spins are relatively weakly coupled
to the environment, and there is a long history of development  of 
experimental
techniques for manipulating the spins using radio frequency pulses.  

A number of groups have already demonstrated the use of NMR computers
\cite{Cory97a,Cory97b,Gershenfeld97,Chuang97,Chuang98a,Chuang98b,
Jones98a,Jones98b,Cory98}.
One
of the key challenges is to try to increase the size of the system used.  
Previous
work on implementing quantum algorithms has focussed on two algorithms in
particular, the Deutsch-Jozsa \cite{Deutsch92} algorithm for distinguishing 
between balanced and
constant functions and Grover's algorithm \cite{Grover97} for searching a 
database. Previous
work on both of these algorithms has used NMR computers with two qubits. In 
this
paper we take the study further by implementing  the Deutsch-Jozsa  
algorithm
for a system of three qubits. A particularly notable feature of the 
experiments
we describe is the use of {\em simultaneous line-selective pulses} to implement 
the key 
stage
of the algorithm,  quantum gates which are closely related to the
controlled-controlled-not gate. 

The  Deutsch-Jozsa algorithm which we will implement is to distinguish
between two classes of two-bit binary functions:
\begin{eqnarray}
f:\{ 0,1 \} \times \{ 0,1 \}\mapsto \{ 0,1 \}.
\end{eqnarray}
The two classes are the {\em constant} functions, in which all input values 
get
mapped to the same output value, and the {\em balanced} functions in which
exactly two of the inputs get mapped to $0$.   The eight balanced or constant
functions are given in Table \ref{table-function1}.

The point of the Deutsch-Jozsa algorithm is that it is possible to decide 
whether a function is 
constant or balanced with only one evaluation of the
function $f$.

The theoretical steps of the quantum algorithm are as follows: 

{\bf [1]} {\bf Preparation:}
Prepare the system in the (pure) state
$
\psi_1 = |0\rangle  |0\rangle |0\rangle
$. 

{\bf [2]} {\bf Excitation:} 
Perform rotations of the spins
about the $y$-axis so that the state becomes 
$
\psi_2 = (|0\rangle+|1\rangle)(|0\rangle+|1\rangle)(|0\rangle-|1\rangle)
$.

{\bf [3]} {\bf Evaluation:}
This is done by implementing the unitary transformation
\begin{eqnarray}
|i\rangle |j\rangle |k\rangle \mapsto |i\rangle |j\rangle |k + f(i,j)\rangle,
\end{eqnarray}
where the addition is performed modulo two.  
The three qubits are now in the state
\begin{eqnarray}
\sum_{i,j=1}^2 (-1)^{f(i,j)}|i\rangle |j\rangle (|0\rangle-|1\rangle).
\end{eqnarray} 
For example in the case of the function $f_4$, the state is
$
\psi_3 = -(|0\rangle-|1\rangle)(|0\rangle+|1\rangle)(|0\rangle-|1\rangle).
$ 

The function $f_4$
is implemented by applying the unitary operator
\begin {eqnarray}
U_4=\left(\begin{array}{cccccccc}
0& 1& 0& 0& 0& 0& 0& 0\\
1& 0& 0& 0& 0& 0& 0& 0\\
0& 0& 0& 1& 0& 0& 0& 0\\
0& 0& 1& 0& 0& 0& 0& 0\\
0& 0& 0& 0& 1& 0& 0& 0\\
0& 0& 0& 0& 0& 1& 0& 0\\
0& 0& 0& 0& 0& 0& 1& 0\\
0& 0& 0& 0& 0& 0& 0& 1\\
\end{array}\right)
\end{eqnarray}
to the state.  We note that this may be written as
\begin {eqnarray}
U_4=\left(\begin{array}{cccc}
2\sigma_x& 0& 0& 0\\
0&2\sigma_x& 0& 0\\
0& 0& E& 0\\
0& 0& 0&E\\
\end{array}\right)
\end{eqnarray}
where $\sigma_x$ is the Pauli matrix normalized so that ${\rm 
tr}(\sigma_x^2)={1\over 2}$
and $E$ is the $2\times 2$ identity matrix.  For convenience we will denote 
such block diagonal matrices by the symbol $\Delta$, so that we write
\begin {eqnarray}
U_4=\Delta(2\sigma_x,2\sigma_x,E,E).
\end{eqnarray} 
The complete list of unitary operators corresponding to the eight balanced or 
constant functions is given in Table \ref{table-unitaries}.

{\bf [4] }{\bf Observation:}
By rotating back by the inverse of the transformation applied in the 
excitation stage, it may be noted that the outputs from the unitary operators 
corresponding to the constant functions $f_1$ and $f_2$ have a component 
proportional to $ |0\rangle |0\rangle |0\rangle$, whereas the outputs from the 
balanced functions $f_3 \ldots f_8$ are orthogonal to this vector so that the 
constant and balanced functions may be distinguished from each other with 
probability one by a von Neumann measurement.

\medskip
The ensemble nature of an NMR quantum computer means that the implementation 
of the algorithm differs somewhat from the theoretical version.  In particular 
the preparation  stage differs since the states of the system are not pure
and the observation stage does not use  a von Neumann 
measurement but measures the amplitudes of  spectral lines.  Nonetheless 
the key goal of the algorithm remains to determine whether  the  
unitary operator   which acts on the system  in the third step corresponds 
to a constant or balanced function.

One way to proceed would be to follow \cite{Cory97b} and produce a pseudo-pure 
state in 
the preparation stage.  In terms of product operators, the state corresponding 
to the pure state 
$
\psi_1 = |0\rangle  |0\rangle |0\rangle
$
is written
\begin{eqnarray}
\rho(\psi_1) = \I_z + \S_z + \R_z + 2\I_z\S_z + 2\I_z\R_z + 2\S_z\R_z + 
4\I_z\S_z\R_z.
\end{eqnarray}
$\I$ refers to the first spin, $\S$, the second and $\R$, the third.

This state is excited to
$
\rho(\psi_2) = \I_x + \S_x - \R_x + 2\I_x\S_x - 2\I_x\R_x - 2\S_x\R_x - 
4\I_x\S_x\R_x
$
in stage {\bf [2]}.

In the evaluation stage, the state to which the spins evolve  depends on which 
function is being implemented.  For example the unitary operator corresponding 
to $f_4$ produces
$
\rho(\psi_3) = - \I_x + \S_x - \R_x  -2\I_x\S_x + 2\I_x\R_x - 2\S_x\R_x + 
4\I_x\S_x\R_x.
$
The full list of output states is given in Table \ref{pure-outputs}.

It should be noted that, of the observable terms (i.e. those terms   
linear
in $\I_x$, 
$\S_x$ and  $\R_x$), the term in $\R_x$ always has the same phase, but 
 the  balanced functions have altered signs of 
$\I_x$ or $\S_x$, or both.
Thus if one observes that the $\I_x$ or $\S_x$ (or both) quartets are inverted
one knows that the function is balanced.

We note however that the same goal can be achieved by starting with thermal 
rather than pure 
initial states.  This is because, as we will show below,  similar effects are 
observed from the outputs starting with thermal initial  states as were 
visible 
starting from pure initial states.  This is not the first time that it has 
been 
noted that in NMR quantum computers, thermal initial states are sufficient to 
implement the algorithms of interest \cite{Chuang98a}.

Thus in the NMR implementation that we will use, the theoretical steps 
{\bf[1]} 
to {\bf [4]} are replaced with
\medskip

{\bf [1${}^\star$]} {\bf Preparation:} 
One starts with the thermal initial state
\begin{eqnarray}
\I_z + \S_z + \R_z 
\end{eqnarray}

{\bf [2${}^\star$]} {\bf Excitation:}
Apply a hard $\pi/ 2$ pulse along the $y$-axis to 
arrive at 
\begin{eqnarray}
\I_x + \S_x + \R_x
\end{eqnarray}

{\bf [3${}^\star$]} {\bf Evaluation:} 
Now evolve the system with one of the unitary operators given in Table 
\ref{table-unitaries}.  This is achieved by using simultaneous line selective 
pulses (see below).  

For example under $f_4$ the state evolves to
\begin{eqnarray}
2\I_x\R_x + \S_x + \R_x.
\end{eqnarray}
The  list of states to which each of $\I_x,\  \S_x$ and $ \R_x$ evolve is 
given in  Table 
\ref{table-theoretical-observations}.

However (see below) the line selective pulses produce evolution by a
unitary  operator which is  close to that required but differs by a controlled
phase  shift. For example, in the case of $f_4$, the line selective pulse
produces the  unitary transformation
\begin{eqnarray}
\Delta(2i\sigma_x,2i\sigma_x,E,E),
\end{eqnarray}
whereas the unitary given in Table \ref{table-unitaries} is
\begin{eqnarray}
U_4 =\Delta(2\sigma_x,2\sigma_x,E,E).
\end{eqnarray}
The relation between these two matrices is 
\begin{eqnarray}
U_4 =\Delta(2\sigma_x,2\sigma_x,E,E) = \Delta(2i\sigma_x,2i\sigma_x,E,E)
\times \Delta(-iE,-iE,E,E).
\end{eqnarray}
The second matrix on the right hand side of this equation is a 
$z$ rotation on the first spin by the angle $\pi / 2$. Thus if one wants
to implement
$U_4$, it would be necessary to follow the line-selective pulse by a phase 
shift.  One finds that similar phase shifts are required for all functions 
except 
$f_1$ and $f_2$.

{\bf [4${}^\star$]} {\bf Observation:}
Under evolution by 
the unitary operators corresponding to any of the balanced functions, either 
the $I$ response or the $S$ response (or both) disappears.  Had we started with 
a pure initial 
state the equivalent line would have been inverted.  

We note that the disappearance or otherwise of the $I$ or $S$ response
is not affected by the final phase shift.  This is because the state
\begin{eqnarray}
{\mathbf I}_x +  {\mathbf S}_x + {\mathbf R}_x
\end{eqnarray}
still evolves to states in which the same line disappears even if this last 
phase shift is not implemented.  This may be appreciated by looking at the 
product 
operators to which the state evolves,
as given in Table \ref{table-experimental-observations}.

\section{Experimental Realization}

One possible way to implement the  evaluation stage of the algorithm
would be to make use of the fact \cite{Barenco95} that any unitary 
transformation can be built up from combinations of the {\em controlled not}
operation and operations on a single qubit. 
The implementation of a controlled not operation by 
magnetic resonance involves the preparation of nuclear magnetization vectors  
of a given spin aligned in opposite directions in the transverse plane.   This 
``anti-phase'' 
condition, which may be represented in the product operator formalism as (say) 
$2\I_y\S_z$, can be generated in a coupled two-spin system through the initial 
stages of 
the INEPT pulse sequence \cite{Morris79}, relying on (refocused) evolution 
under the $2\I_z\S_z$  
operator for a fixed interval $1/(2J_{IS})$.   However, the extension of this 
procedure 
to more than two coupled spins is complicated and not easy to implement.    
A more direct approach, and the one we have employed, is through the use of 
high-selectivity radio-frequency 
pulses designed to perturb transverse magnetization one line at a time.     For 
example applying a $\pi$ pulse with 
Hamiltonian of the form \cite{Sorensen83}
\begin{eqnarray}
\R_x +2\I_z \R_x + 2\S_z\R_x + 4\I_z\S_z \R_x
\end{eqnarray} 
causes the system to evolve by the unitary operator
\begin{eqnarray}
\Delta(2i\sigma_x,E,E,E).\label{single-pulse}
\end{eqnarray} 
The key observation from the point of view of our work is that more than one 
such line-selective perturbation may be applied simultaneously \cite{Kupce93}.  
 Thus any of the unitary operators in Table 
\ref{table-experimental-observations} (and indeed a very wide class of 
controlled rotations about more general axes)
may be 
produced in the same time that is required to
produce the perturbation given in (\ref{single-pulse}).  It is worth noting 
that this time is of the 
same order as that required to implement the INEPT sequence. We feel that as 
well as being helpful for
the present work, the method of manipulating spins via simultaneous line 
selective pulses 
may well
prove advantageous in NMR quantum computers with more spins.

	The experimental task is to shape the radio-frequency pulse envelope 
so as to achieve 
sufficient selectivity in the frequency domain that there is negligible 
perturbation 
of the next-nearest neighbour of the spin multiplet.   In this sense the 
technique 
resembles that used in pseudo-two-dimensional spectroscopy \cite{Davies87} 
where the 
frequency of a soft radio-frequency pulse is stepped through the spectrum of 
interest in very small frequency increments, exciting the transitions one by 
one.  
We investigated several possible pulse shapes for this purpose, including 
rectangular, Gaussian, sine-bell, and triangular, before settling on the 
Gaussian as the most suitable for the task.  

In a weakly-coupled three-spin $ISR$ system the $R$ spectrum is a 
doublet of 
doublets with splittings $J_{IR}$ and $J_{SR}$.  Application of $\pi$š pulses 
to all four 
transverse $R$-spin magnetization components corresponds to a constant
function in the sense of the Deutsch-Jozsa algorithm, and the ``do 
nothing'' 
experiment represents the other constant function.   The balanced 
functions 
may be implemented by application of soft $\pi$š pulses to the individual 
lines two 
at a time, for example $[0š, 0š, \piš, \piš]$,\ $[0š, \piš, 0š, \piš]$, or 
$[0š, \piš, 
\piš, 0š]$, where $0$š denotes no soft pulse.  These cases, corresponding to
functions $f_3,f_6$ and $f_8$ have Hamiltonians proportional to $\R_x -2\I_z 
\R_x$,  $\R_x -2\S_z \R_x$
and $\R_x -4\I_z \S_z\R_x$ respectively.    One way to calculate the effect of 
these Hamiltonians is to use standard product operator manipulations 
\cite{Sorensen83}. For example one finds that a $\pi$ pulse with  Hamiltonian 
of the 
form $\R_x -2\I_z \R_x$ leaves $\R_x$ and $\S_x$ unchanged and changes $\I_x$ 
to $2\I_y\R_x$ as in Table \ref{table-experimental-observations}.

The practical implementation is deceptively 
simple.  
Starting with a thermal state, a hard $\pi/2$  pulse about the $y$ axis 
(denoted 
$[\pi/2]_y$) excites transverse magnetizations $\I_x$, $\S_x$  and $\R_x$.  The 
evaluation 
step is the application of line-selective $[\pi]_x$ pulses to the 
individual 
components of the $R$ multiplet. We may choose to apply soft $[\pi]_x$  pulses 
to all four magnetization 
components, any two of the four, or none at all.  In all cases the soft pulses 
are 
applied simultaneously, while the remaining transitions are simply left to 
evolve 
freely for the same period of time.  However the perturbed magnetization 
components lose intensity only through spin-spin relaxation during the 
relatively 
long interval of the soft pulse, $T$, because the effects of spatial 
inhomogeneity of the magnetic 
field 
are refocused, whereas the freely precessing components decay more rapidly, 
with a shorter time constant $T_2^*$.  This difference in intensities serves to 
confirm 
which $R$ transitions were perturbed.

Experiments were carried out at 400 MHz on a Varian VXR-400
spectrometer equipped with a waveform generator which controlled the shaped
radiofrequency pulses.     The three-spin proton system chosen for study was
2,3-dibromopropanoic acid in $\hbox{\rm CDCl}_3$.  The three coupling constants 
are 
$J_{IR} =+11.3$  Hz, $J_{IS} = -10.1$ Hz, and $J_{RS} = +4.3$ Hz. (The negative 
sign of the geminal
coupling $J_{IS}$ \cite{Freeman62} has no particular significance in these 
experiments.)   Strong
coupling effects are evident between spins $I$ and $S$, with 
$J_{IS}/\delta_{IS} = 0.12$.

Each soft $[\pi]_x$ pulse 
can be thought of as acting on one of the four $R$-spin magnetizations in a 
rotating 
frame at the exact resonance frequency of that particular $R$ line.  
These four reference 
frames rotate at four different frequencies $ (\pm J_{SR} \pm J_{IR})/2$ with 
respect to the transmitter frequency centred on the $R$ chemical shift.  
The $x$-axes of all four frames must be coincident at the beginning of 
the soft pulse interval $T$.  The duration of the soft pulse, $T$,  may be 
chosen in such a 
way as to optimize the frequency selectivity.

The predicted result (Table \ref{table-experimental-observations}), is to 
convert $I$- or $S$-spin 
magnetization 
into various forms of multiple-quantum coherence in the six cases where the 
$R$ 
magnetization components are perturbed in pairs (the balanced functions) but 
to 
leave the $I$- and $S$-spin magnetizations unaffected in the remaining two 
cases 
where the four $R$ magnetization components are all perturbed or all left 
alone 
(the constant functions).   These predictions are clearly borne out by the 
experimental specta shown in the Figure.  In principle, complete conversion 
into unobservable 
multiple-quantum coherence would be detected by the disappearance of the 
appropriate $I$ or $S$-spin response. In practice, owing to non-idealities of 
the 
system (for example strong coupling effects between $I$ and $S$) this is 
observed 
as a roughly eightfold loss of intensity rather than complete suppression.   

Eight experiments were performed to test the eight cases of Table 
\ref{table-experimental-observations}.   The
transmitter frequency was centred on the $R$-spin multiplet.  Note that the 
$R$
spectrum remains unperturbed throughout the series, except for the intensity
perturbation mentioned above, a result of the refocusing effect of the soft 
$\pi$
pulses.   The phases of the $I$- and $S$ signals will be determined by the
scalar coupling and chemical shift evolution during the period $T$.
These complex phase patterns do not interfere with the 
Deutsch-Jozsa test because this involves only the observation of the 
``disappearance'' 
of
certain signals.   These signal losses are made clearly evident by displaying
absolute-value spectra, which may then be integrated.   The integrated 
intensities
are shown as percentages of the corresponding intensities in the top spectrum 
(no
soft-pulse perturbation).   Creation of multiple-quantum coherence is indicated
by the roughly eightfold decrease in intensity in the appropriate places; all 
other
$I$- and $S$-spin intensities remain essentially at 100\%.   This 
interpretation was
confirmed in a second experiment with a multiplet-selective soft $\pi/2$ pulse
applied to the $R$ spins at the end of the sequence.  This has the effect of 
restoring
the ``lost'' intensities by reconverting $IR$ and $SR$ multiple-quantum 
coherence 
into
observable magnetization.

 Thus, in a single measurement, a distinction can be made between constant
and balanced functions simply on the grounds of the ``disappearance'' of $I$- 
or $S$-
spin lines.   The fact that further details can be gleaned about the pattern of 
soft-
pulse perturbation is irrelevant to the Deutsch-Jozsa algorithm.  The extension 
to
systems of more than three coupled spins is clear.  Because the soft pulses are
applied simultaneously, this involves no increase in the duration of the
perturbation stage.  The main limitation would be the magnitude of the smallest
coupling constant, for this sets the frequency selectivity requirement.  
Extension
to more qubits would most likely invoke the introduction of heteronuclear spins
such as ${}^{13}\hbox{\rm C}$ and ${}^{19}\hbox{\rm F}$.

\bigskip
\noindent{\large\bf Acknowledgments}
The authors are indebted to Dr \={E}riks Kup\v{c}e of Varian
Associates for invaluable advice on the generation of shaped selective
radio-frequency pulses.

} %% closing bracket for openup

\vfill\eject

\begin{table}[p]
\begin{center}
\begin{tabular}{|c|c|c|c|c|c|c|c|c|}
\hline
$x$&$f_1(x)$&$f_2(x)$&$f_3(x)$&$f_4(x)$ 
&$f_5(x)$&$f_6(x)$&$f_7(x)$&$f_8(x)$\\\hline
00& 0 & 1 & 0 & 1& 1 & 0 & 1 & 0\\
01& 0 & 1 & 0 & 1& 0 & 1 & 0 & 1\\
10& 0 & 1 & 1 & 0& 1 & 0 & 0 & 1\\
11& 0 & 1 & 1 & 0& 0 & 1 & 1 & 0\\\hline
\end{tabular}
\end{center}
\caption{The eight possible balanced or constant binary 
functions mapping two bits to one bit.}
\label{table-function1}
\end{table}

\begin{table}[p]
\begin{center}
\begin{tabular}{|l|l|}
\hline
$f $&$U$ \\\hline
$f_1$&$\Delta(E,E,E,E)$\\
$f_2$&$ \Delta(2\sigma_x,2\sigma_x,2\sigma_x,2\sigma_x)$\\
$f_3$&$ \Delta(E,E,2\sigma_x,2\sigma_x)$\\
$f_4$&$\Delta (2\sigma_x,2\sigma_x,E,E)$\\
$f_5$&$ \Delta(2\sigma_x,E,2\sigma_x,E)$\\
$f_6$&$ \Delta(E,2\sigma_x,E,2\sigma_x)$\\
$f_7$&$ \Delta(2\sigma_x,E,E,2\sigma_x)$\\
$f_8$&$ \Delta(E,2\sigma_x,2\sigma_x,E)$\\ \hline
\end{tabular}
\end{center}
\caption{The unitary operators corresponding to the eight  constant or 
balanced
binary functions mapping two  bits to one bit.}
\label{table-unitaries}
\end{table}

\begin{table}[p]
\begin{center}
\begin{tabular}{|l|c|}
\hline
$f $&output \\\hline
$f_1$&$ +\I_x + \S_x - \R_x + 2\I_x\S_x - 2\I_x\R_x - 2\S_x\R_x - 
4\I_x\S_x\R_x$  \\
$f_2$&$ + \I_x + \S_x - \R_x + 2\I_x\S_x - 2\I_x\R_x - 2\S_x\R_x - 
4\I_x\S_x\R_x$ \\
$f_3$&$ - \I_x + \S_x - \R_x  -2\I_x\S_x + 2\I_x\R_x - 2\S_x\R_x +
4\I_x\S_x\R_x$ \\
$f_4$&$ - \I_x + \S_x - \R_x  -2\I_x\S_x + 2\I_x\R_x - 2\S_x\R_x +
4\I_x\S_x\R_x$ \\
$f_5$&$ + \I_x - \S_x - \R_x  -2\I_x\S_x - 2\I_x\R_x + 2\S_x\R_x + 
4\I_x\S_x\R_x$ \\
$f_6$&$ + \I_x - \S_x - \R_x  -2\I_x\S_x - 2\I_x\R_x + 2\S_x\R_x + 
4\I_x\S_x\R_x$ \\
$f_7$&$ - \I_x - \S_x - \R_x + 2\I_x\S_x + 2\I_x\R_x + 2\S_x\R_x - 
4\I_x\S_x\R_x$ \\
$f_8$&$ - \I_x - \S_x - \R_x + 2\I_x\S_x + 2\I_x\R_x + 2\S_x\R_x - 
4\I_x\S_x\R_x$ \\
\hline
\end{tabular}
\end{center}
\caption{The output states from a (pseudo) pure initial state after the 
evaluation stage. 
}
\label{pure-outputs}
\end{table}

\begin{table}[p]
\begin{center}
\begin{tabular}{|l|l|c|c|c|}
\hline
$f $&$U$&${\mathbf I}_x$&${\mathbf S}_x$&${\mathbf R}_x$ \\\hline
$f_1$&$\Delta(E,E,E,E)$&${\mathbf I}_x$&${\mathbf S}_x$&${\mathbf 
R}_x$\\
$f_2$&$ \Delta(2\sigma_x,2\sigma_x,2\sigma_x,2\sigma_x)$&${\mathbf I}_x$
	&${\mathbf S}_x$&${\mathbf R}_x$\\
$f_3$&$ \Delta(E,E,2\sigma_x,2\sigma_x)$
	&$2{\mathbf I}_x{\mathbf R}_x$&${\mathbf S}_x$&${\mathbf R}_x$\\
$f_4$&$ \Delta(2\sigma_x,2\sigma_x,E,E)$
	&$2{\mathbf I}_x{\mathbf R}_x$&${\mathbf S}_x$&${\mathbf R}_x$\\
$f_5$&$ \Delta(2\sigma_x,E,2\sigma_x,E)$
	&${\mathbf I}_x$&$2{\mathbf S}_x{\mathbf R}_x$&${\mathbf R}_x$\\
$f_6$&$ \Delta(E,2\sigma_x,E,2\sigma_x)$
	&${\mathbf I}_x$&$2{\mathbf S}_x{\mathbf R}_x$&${\mathbf R}_x$\\
$f_7$&$ \Delta(2\sigma_x,E,E,2\sigma_x)$
	&$2{\mathbf I}_x{\mathbf R}_x$&$2{\mathbf S}_x{\mathbf R}_x$&${\mathbf 
R}_x$\\
$f_8$&$ \Delta(E,2\sigma_x,2\sigma_x,E)$ 
	&$2{\mathbf I}_x{\mathbf R}_x$&$2{\mathbf S}_x{\mathbf R}_x$&${\mathbf 
R}_x$\\
\hline
\end{tabular}
\end{center}
\caption{The effect on input product operators 
${\mathbf I}_x,\ {\mathbf S}_x$ and ${\mathbf R}_x$ of the unitary operators 
in the 
second column; for example under $\Delta(E,2\sigma_x,E,2\sigma_x)$, the 
input 
${\mathbf S}_x$ evolves to $2{\mathbf S}_x{\mathbf R}_x$}
\label{table-theoretical-observations}
\end{table}

\begin{table}[p]
\begin{center}
\begin{tabular}{|l|l|c|c|c|}
\hline
$f $&$U$&${\mathbf I}_x$&${\mathbf S}_x$&${\mathbf R}_x$ \\\hline
$f_1$&$\Delta(E,E,E,E)$&${\mathbf I}_x$&${\mathbf S}_x$&${\mathbf 
R}_x$\\
$f_2$&$ \Delta(2i\sigma_x,2i\sigma_x,2i\sigma_x,2i\sigma_x)$&${\mathbf 
I}_x$
	&${\mathbf S}_x$&${\mathbf R}_x$\\
$f_3$&$ \Delta(E,E,2i\sigma_x,2i\sigma_x)$
	&$2{\mathbf I}_y{\mathbf R}_x$&${\mathbf S}_x$&${\mathbf R}_x$\\
$f_4$&$ \Delta(2i\sigma_x,2i\sigma_x,E,E)$
	&$-2{\mathbf I}_y{\mathbf R}_x$&${\mathbf S}_x$&${\mathbf R}_x$\\
$f_5$&$ \Delta(2i\sigma_x,E,2i\sigma_x,E)$
	&${\mathbf I}_x$&$-2{\mathbf S}_y{\mathbf R}_x$&${\mathbf R}_x$\\
$f_6$&$ \Delta(E,2i\sigma_x,E,2i\sigma_x)$
	&${\mathbf I}_x$&$2{\mathbf S}_y{\mathbf R}_x$&${\mathbf R}_x$\\
$f_7$&$ \Delta(2i\sigma_x,E,E,2i\sigma_x)$
	&$-4{\mathbf I}_y{\mathbf S}_z{\mathbf R}_x$&$-4{\mathbf I}_z{\mathbf 
S}_y{\mathbf R}_x$&${\mathbf R}_x$\\
$f_8$&$ \Delta(E,2i\sigma_x,2i\sigma_x,E)$ 
	&$4{\mathbf I}_y{\mathbf S}_z{\mathbf R}_x$&$4{\mathbf I}_z{\mathbf 
S}_y{\mathbf 
R}_x$&${\mathbf R}_x$\\
\hline
\end{tabular}
\end{center}
\caption{The effect on input product operators 
${\mathbf I}_x,\ {\mathbf S}_x$ and ${\mathbf R}_x$ of the unitary operators 
in the 
second column; for example under $\Delta(E,2i\sigma_x,E,2i\sigma_x)$, 
the input 
${\mathbf S}_x$ evolves to $2{\mathbf S}_y{\mathbf R}_x$.}
\label{table-experimental-observations}
\end{table}

\begin{figure}[bt]
\begin{picture}(100,400)(100,330) \put(10,0){
\scalebox{1.00}[1.00]{\psfig{file=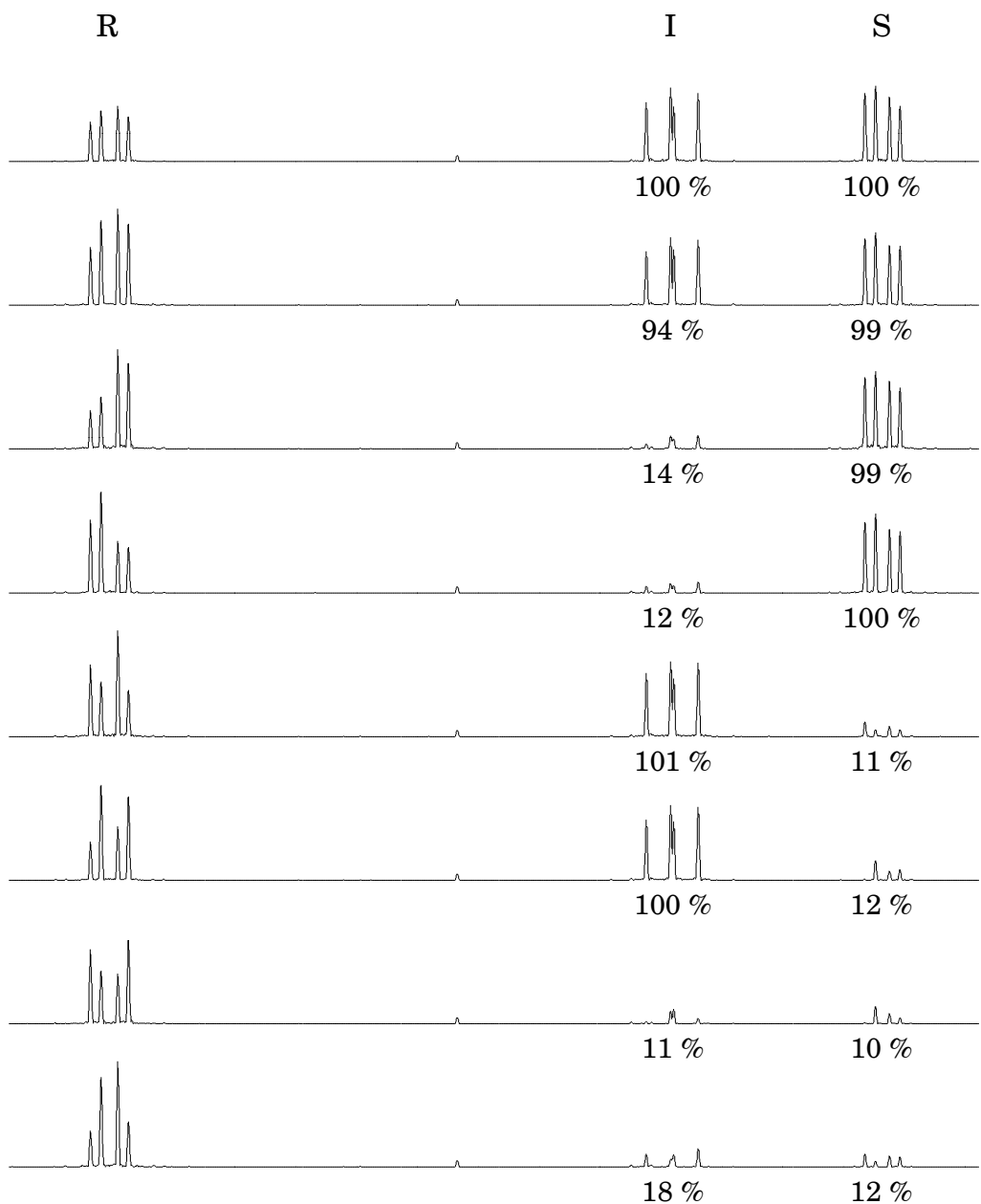}}
} \end{picture}
\caption{
Eight absolute-value 400 MHz spectra of 2,3-dibromopropanoic acid
obtained with the eight different perturbations set out in Table 
\ref{table-experimental-observations}. The soft pulses
were applied simultaneously with a pulse duration $T = 0.65$ seconds.  Reading
from top to bottom, these spectra correspond to the functions $f_1 . . . f_8$ 
of Table
\ref{table-function1}.  Integrals of the $I$- and $S$-spin responses are shown 
as percentages of those in
the top trace.  After the evaluation of these integrals, the line shapes were
improved by pseudo-echo weighting.  Note the suppression of the appropriate 
$I$-
and $S$-spin responses by about an order of magnitude.
}
\label{figure}
\end{figure}

\end{document}